# Inherent Limit to Coherent Phonon Generation under Non-Resonant Light Field Driving

K. Uchida[1], K. Nagai[1], N. Yoshikawa[1], and K. Tanaka[1,*]

[1] *Department of Physics, Graduate School of Science, Kyoto University, Kyoto, Kyoto 606-8502, Japan*

**Abstract**

Coherent manipulation of quasi-particles is a crucial method to realize ultrafast switching of the relating macroscopic order. In this letter, we studied coherent phonon generation under strong light field which allows to induce non-perturbative nonlinear optical phenomena in solids. The efficiency of coherent phonon generation starts to saturate and deviate from the pronounced linear power dependence when the light intensity goes into the non-perturbative regime. We propose a novel theoretical model based on Floquet picture and show that the saturation is due to a limitation of the driving force inherent in non-resonant driving of the electronic system in the non-perturbative regime.

**Main text**

Intense laser field driving of solid materials may give rise to a novel non-equilibrium state that has the potential to show new functionality and induce phase transition phenomena [1,2]. Such a state can be realized through a large-amplitude excitation of a phonon system, and this capability has given rise to the study of nonlinear phononics [3,4]. One promising driving method for the phonon mode is impulsive stimulated Raman scattering (ISRS) process [5,6], where ultrashort-pulse excitation in a non-resonant relation to the electronic transition generates an impulsive force through electron-phonon coupling. In the perturbation limit, ISRS gives a driving force proportional to the light intensity [5,6]. It is a crucial problem for the realization of nonlinear phononics how a large-amplitude phonon mode can be realized through electron-phonon coupling using extremely strong light beyond the range in which perturbation theory applies (non-perturbative nonlinear regime) [7].

The recent development of femtosecond laser technology in the mid-infrared (MIR) region enables us to step into an non-perturbative nonlinear regime in solids without damaging the sample. In this regime, the electronic energy is significantly modified, and highly nonlinear optical responses such as high-order harmonic generation (HHG) appears [8-12]. Thus, this technology has made it possible to conduct experiments aimed at answering the question of whether there is a limit to the driving force of coherent phonon generation under non-resonant conditions.

Here, we study coherent phonon generation induced by using intense MIR laser field ($\hbar\Omega_{MIR} = 0.26$ eV), which drives an electronic system strongly. The maximum intensity at the sample position was 0.6 TW/cm$^2$ in air, corresponding to an electric field of $E_{MIR} = 13$ MV/cm inside the sample. The sample was a single crystal of 100 μm-thick gallium selenide (GaSe), which shows efficient HHG emission, i.e., strong light-electron interactions [13-15]. Since the photon energy of the MIR pulse (0.26 eV) is much larger than the phonon energy in GaSe (< 0.04 eV) [16], direct excitation of phonon modes is sufficiently suppressed. Figure 1(a) shows the experimental setup of the HHG measurement in transmission geometry. We performed experiments with MIR polarizations along the armchair (A-pol.) and zigzag (Z-pol.) directions. The measured HHG spectra are shown in Fig. 1b. We observed HHG from 5th to 12th orders and found that HHG with A-pol. is stronger than that with Z-pol., which is consistent with previous research [15]. Figure 1c shows the MIR intensity dependence of 5th-order harmonic intensity. Above 0.3 TW/cm$^2$, the plot starts to deviate from a power law ($I_{HHG} \propto I_{MIR}^5$), indicating that the light-electron interaction enters the non-perturbative nonlinear regime.

Under such an extremely nonlinear condition, we performed MIR-pump and near-infrared (NIR)-probe ($\hbar\Omega_{NIR} = 1.55$ eV) measurements, as depicted in Fig. 2a, to observe coherent phonons in GaSe [6,17]. The transmitted NIR probe pulses were detected in a polarization-resolved manner, which is a similar technique to the electro-optic sampling method (see Sec. 2 of Supplementary Information for detail). Figure 2b plots the pump-probe signal as a function of the time delay T between MIR pump and NIR probe pulses. After the impulsive response at T = 0 ps, which may be attributed to a third order or higher nonlinear coherent optical process, two kinds of oscillations with different frequencies appear clearly. The faster oscillation decays within 10 ps, while the slower one continues to oscillate for least 100 ps. Figure 2c shows the Fourier-transformed spectrum of the signal, in which there are two distinct peaks at frequencies of 0.6 THz and 4.0 THz corresponding to the temporal oscillations. These frequencies are respectively attributed to the $E_{2g}$ shearing mode and $A_{1g}$ breathing mode depicted in Fig. 2d [18].

To confirm the properties of the coherent phonons, we performed a fitting of temporal profiles by using the following equation:

$$I(\mathrm{T}) = \sum_{i=A_{1g}, E_{2g}} A_i \exp(-T/\tau_i)\cos(2\pi f_i T + \phi_i). \qquad (1)$$

Here, $A_i$, $\tau_i$, $f_i$, and $\phi_i$ are respectively the amplitude, decay time, frequency, and initial phase of the $i$-th ( = $E_{2g}$ and $A_{1g}$ ) mode. Figures 3 a-f show the MIR intensity dependence of the coherent phonon properties (decay time, phase, and frequency) of the $E_{2g}$ and $A_{1g}$ modes. The properties are independent of MIR intensity for both modes, and the frequencies are almost the same as those in Raman spectroscopy [18]. This clearly indicates that anharmonicity of phonons was negligible under the experimental conditions, and the phonon dynamics can be described by a simple damped harmonic oscillator model, given by

$$\frac{d^2}{dt^2} Q_i + \frac{2}{\tau_i} \frac{d}{dt} Q_i + \left(2\pi \tilde{f}_i\right)^2 Q_i = F_i(t), \qquad (2)$$

where $Q_i$ and $F_i(t)$ are the normal coordinate and the driving force for the $i$-th mode phonon, respectively, and $\tilde{f}_i$ is given by $(f_i^2 - 1/(2\pi\tau_i)^2)^{1/2}$.

Since the pulse width of MIR electric field $\Delta \tau$ (60 fs in full width half maxima) is shorter than the period of phonon oscillation $1/f_i$ (> 250 fs), driving force can be regarded as an impulsive force ($F_i(t)=F_i\delta(t)$), and the amplitude of phonon is proportional to $F_i$.

In contrast to its simple dynamics, regarding the amplitude of the phonons, both modes show saturation behavior above 0.3 TW/cm$^2$ (Figs. 4a and b). The threshold intensity of saturation depends on the phonon mode and MIR polarization. In the $E_{2g}$ mode, much stronger saturation was observed when the MIR polarization was Z-pol. than when it was A-pol. (Fig. 4a). On the other hand, in the $A_{1g}$ mode, the saturation behavior was almost independent of the MIR polarization (Fig. 4b).

Since the temporal phonon dynamics are well described by Eq. (2), these results indicate that the driving force for phonon $F_i$ shows saturation as the MIR intensity increases. Saturation of the driving force depending on the phonon mode and MIR polarization cannot be explained by the conventional ISRS process in the perturbative regime [5,6], where the driving force is proportional to the MIR intensity and its amplitude is independent of the MIR polarization (see Sec. 2 in Supplemental Information for detail). The fact that saturation threshold intensities are almost the same as those in HHG implies that saturation of the phonon amplitude is also attributed to the non-perturbative light-electron interaction.

To understand the saturation mechanism of the driving force, we can use a simple two-level model with the Born-Oppenheimer approximation to calculate the impulsive force for phonons in the non-perturbative regime. A two-level system linearly coupled with a phonon mode interacting with a light field can be described by the Hamiltonian in the non-perturbative regime, as follows:

$$\hat{H}(t) = \varepsilon_g(Q_i)|g\rangle\langle g| + \varepsilon_e(Q_i)|e\rangle\langle e| + dE_{MIR}(t)|g\rangle\langle e| + d^*E_{MIR}(t)|e\rangle\langle g|. \tag{3}$$

Here, $|g\rangle$ ( $|e\rangle$ ) is the electronic ground (excited) state with energy of $\varepsilon_g(Q_i)$ ( $\varepsilon_e(Q_i)$ ) with respect to a normal coordinate $Q_i$, which works as an adiabatic potential for the $i$-th mode phonon, as shown by the dashed lines in Fig. 4c. $d=\langle g|\hat{d}|e\rangle$ is the transition dipole moment between the ground and excited states, and $E_{MIR}(t)$ is the temporal profile of the MIR electric field.

Without the laser field ($E_{MIR}(t) = 0$), we assume that the system is in the ground state $|g\rangle$ with no oscillation. Therefore, the driving force of the phonon mode results in

$$F_i = -\left.\frac{\partial \varepsilon_g(Q_i)}{\partial Q_i}\right|_{Q_i=0} = 0 . \tag{4}$$

Under an intense laser field, it is hard to neglect the excited state which is mixed with the ground state through the light-electron interaction. One way to solve this problem is the Floquet picture [19,20]. Here, we consider a "dressed" ground (excited) state $|g*\rangle$ ( $|e*\rangle$ ), whose adiabatic potentials are modified depending on the field strength and frequency of the MIR pulse, as shown in Fig. 4c. Mixing in the adiabatic potential of the bare excited state $|e\rangle$ gives rise to a finite impulsive driving force in the "dressed" ground state $|g*\rangle$.

To estimate the impulsive force in the "dressed" ground state, we first assume continuous non-resonant excitation light $E_{MIR}(t)=E_{MIR}\cos(\Omega_{MIR} t)$ ($\Delta\varepsilon(0) = \varepsilon_e(0) - \varepsilon_g(0) \gg \hbar\Omega_{MIR}$) and apply the rotating wave approximation (RWA). With this approximation, we can analytically derive the energy shifts of the driven two-level system in the non-perturbative regime, i.e., Rabi splitting [21], and the driving force as follows:

$$F_i = -\frac{1}{2}\frac{\partial \varepsilon_{g*}(Q_i)}{\partial Q_i}\bigg|_{Q_i=0} = -\frac{1}{2}\frac{\partial \varepsilon_e(Q_i)}{\partial Q_i}\bigg|_{Q_i=0}\left(1 - \frac{1}{\sqrt{\xi^2+1}}\right), \tag{5}$$

$$\xi = \frac{dE_{MIR}}{\varepsilon_e(0) - \varepsilon_g(0) - \hbar\Omega_{MIR}}. \tag{6}$$

Here, $\xi$ is the parameter in nonlinear optics for the perturbation expansion [22].

In general, when $\xi$ becomes comparable to unity, a non-perturbative nonlinear optical response such as HHG can be observed. Figure 4d plots the driving force of phonon $F_i$ as a function of $\xi^2$. When $\xi$ is much smaller than one, $F_i$ is proportional to $\xi^2$ or $(E_{MIR})^2$, which is similar to the prediction in the ISRS process (dashed line). On the other hand, when $\xi$ becomes unity, it starts to saturate. Since $\xi \sim 1$ corresponds to $I_{MIR} \sim 0.3$ TW/cm$^2$ for our experimental conditions, the calculation is in good agreement with our experimental results showing saturation of the driving force in association with HHG.

Moreover, when $\xi$ is much larger than one, the driving force gradually approaches the upper limit given by half the driving force in the bare excited state $F_i = -(\partial \varepsilon_e(Q_i)/\partial Q_i)/2$. In the strong field limit, the ratio of the ground to excited states in the "dressed" ground state becomes unity.

Therefore, the driving force on the phonon in the dressed state is described as the average of the driving forces in the bare ground and excited states.

To evaluate the driving force more exactly, we performed a numerical calculation without the RWA (see Sec.4 in Supplemental Information for detail), taking into account the Bloch-Siegert shift contribution [23], which is neglected in Eq. (5). The result is shown by the blue solid line in Fig. 4d, indicating that the Bloch-Siegert shift contribution introduces a slight increase of the driving force but gives qualitatively the same result as the calculation with the RWA (Eq. (5)). In the limit $\xi \ll 1$, the numerical result completely coincides with the theoretical value of the ISRS process. These results provide us an answer to the initial question that there should be a limitation to the non-resonant coherent phonon generation.

At the moment, we have no explanation for why the saturation threshold intensity depends so strongly on the phonon mode and MIR polarization. However, we suppose that it reflects an electron-phonon coupling depending on electronic bands and crystal momentum which cannot be accessed by conventional Raman spectroscopy.

In conclusion, we observed coherent phonon generation in the extreme nonlinear regime where non-perturbative nonlinear optical responses such as HHG are observed. Although the phonon dynamics can be described in terms of a damped harmonic oscillator, the phonon amplitude showed saturation depending on the crystal orientation and mode. This saturation behavior can be qualitatively explained by a two-level model with the Born-Oppenheimer approximation and suggests that there is an upper limit to the driving force for the phonon mode in the non-resonant laser excitation. This scenario applies not only to phonons but also to other quasiparticles, such as magnons [24], and suggests that we need precise control of driving electric field or alternative means such as direct excitation of quasiparticles in order to realize the formation of macroscopic ordering through the large amplitude driving of quasiparticles.

**Acknowledgements**

This work was supported by a Grant-in-Aid for Scientific Research (S) (Grant No. 17H06124). K. U. was supported by a Grant-in-Aid for Research Activity Start-up (Grant No. 18H05850). The authors are thankful to Yosuke Kayanuma for the fruitful discussion.

**Author contribution**

K. U. and K. T. conceived the experiments. K. U., K. N., and N. Y. set up and carried out the experiments. K. U. and K. T. constructed theoretical framework and performed the calculation. K. U. and K. T. wrote the manuscript. All the authors contributed to the discussion and interpretation of the results.

**Figure captions**

**Figure 1** HHG in GaSe crystal. (a) Setup of HHG experiment in transmission geometry. The MIR polarization along the armchair (zigzag) direction is denoted as the A-(Z-) polarization. (b) HHG spectra in visible region with A-pol. (orange solid line) and Z-pol. (blue solid line). (c) MIR intensity dependence of 5th order HHG intensity with A-pol. (orange circles) and Z-pol. (blue open squares). Dashed lines are guides for the eye that are proportional to the 5th power of MIR intensity.

**Figure 2** Coherent phonon spectroscopy in GaSe. (a) Experimental setup of coherent phonon detection in transmission geometry. The angle between the MIR and NIR polarizations was 45 degrees. The polarization modulation of the transmitted NIR probe pulse was detected by EO sampling. (b) Pump probe signal as a function of time delay with A-pol. and MIR intensity of 0.6 TW/cm$^2$. Two temporal regions from -1 ps to 10 ps and from 15 ps to 150 ps are plotted for clarity. (c) Fourier spectra of the pump-probe signal shown in b. d, Schematics of $E_{2g}$ and $A_{1g}$ phonon modes in e-GaSe. Orange (green) arrows indicate the direction of displacement of the Se (Ga) atom.

**Figure 3** MIR intensity dependence of coherent phonon properties. Fitting results for $E_{2g}$ ($A_{1g}$) mode with Z-pol. are shown in a, c, and e (b, d, and f). Decay time (a and b), phase (c and d) and frequency (e and f) of oscillations are plotted as a function of MIR intensity.

**Figure 4** Saturation of phonon amplitude in extremely nonlinear regime. Amplitude of (a) $E_{2g}$ and (b) $A_{1g}$ phonon modes as a function of MIR intensity. Solid and open squares (circles) are respectively indicate results with A-pol.

and Z-pol. The dashed lines are guides for the eye that is proportional to MIR intensity. (c) Bare (dashed line) and Dressed (solid line) two-level system (g: ground state, e: excited state) depending on phonon coordinate Q in the Frank-Condon picture. Modification of the ground-state energy surface through photon dressing causes a driving force on the phonon. (d) Calculated results of driving force on phonon. Red and blue solid lines respectively show the analytical result with the RWA (Eq. (5)) and the numerical result without the RWA. Green dashed line shows the driving force described by the ISRS process. The dotted line indicates the upper limit of the driving force within RWA.

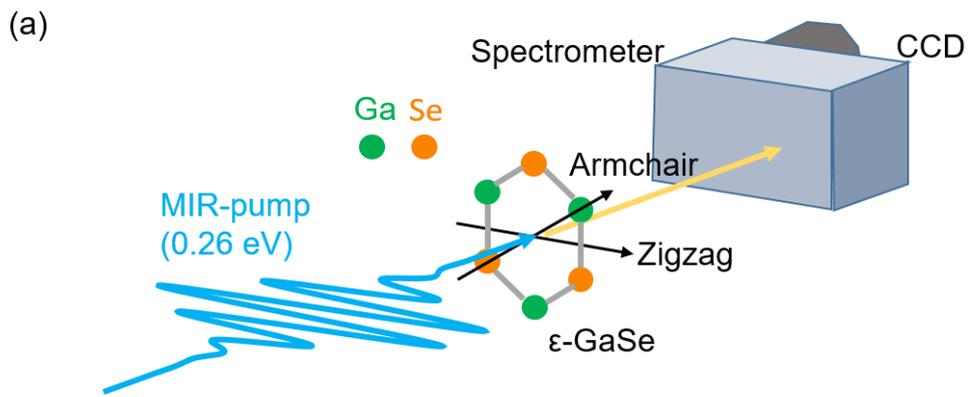

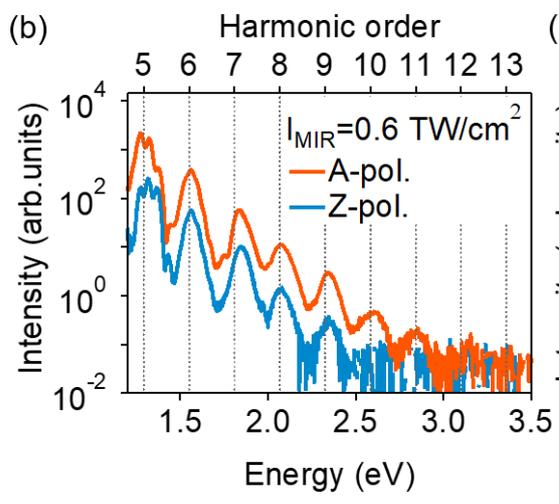
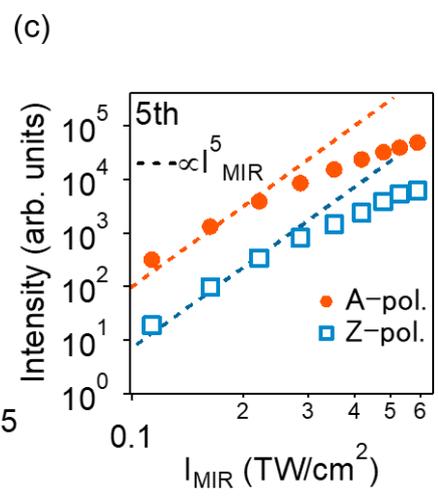

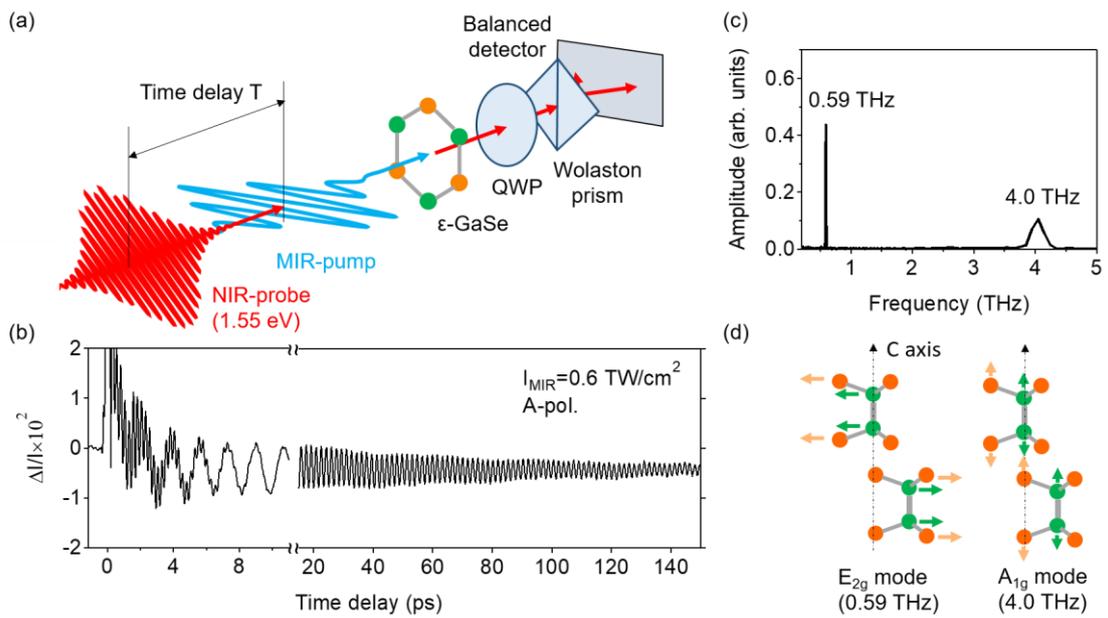

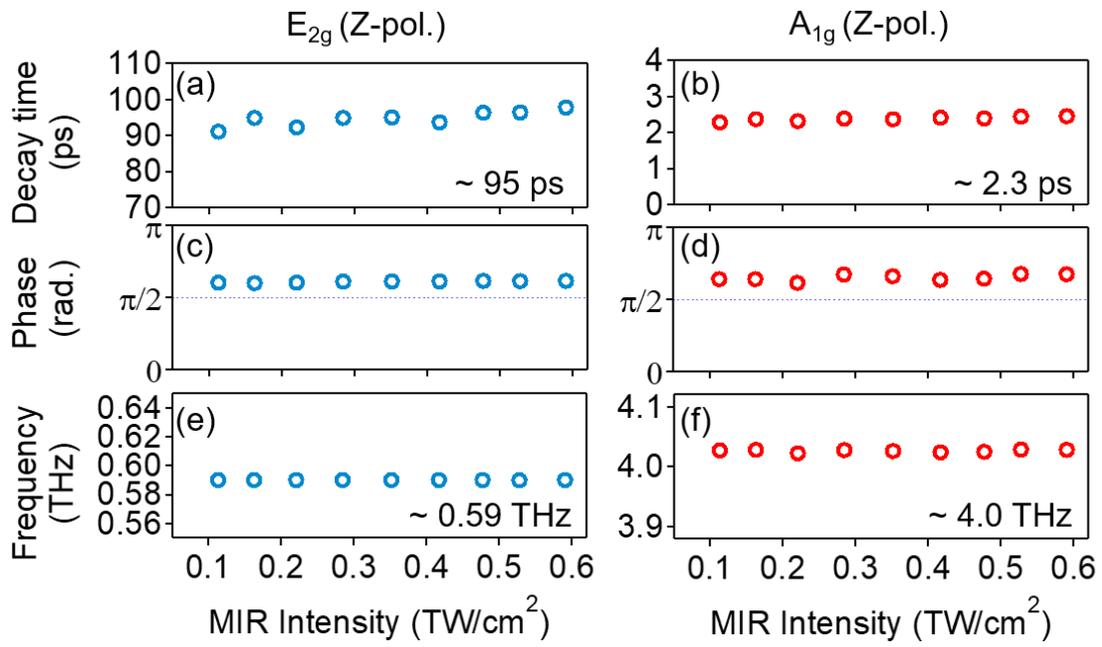

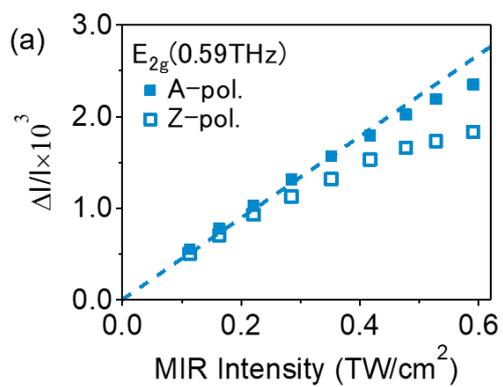
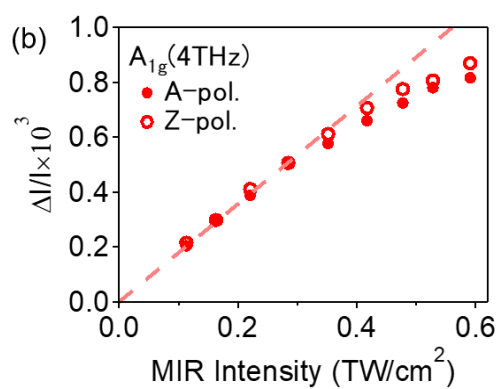
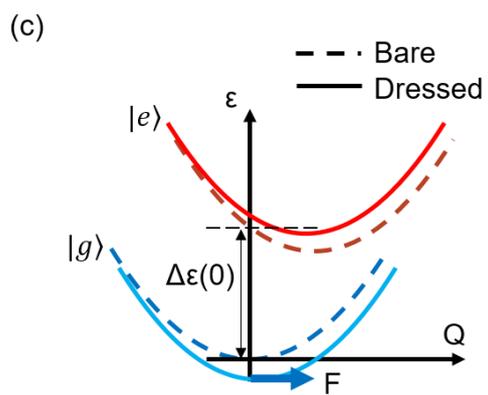
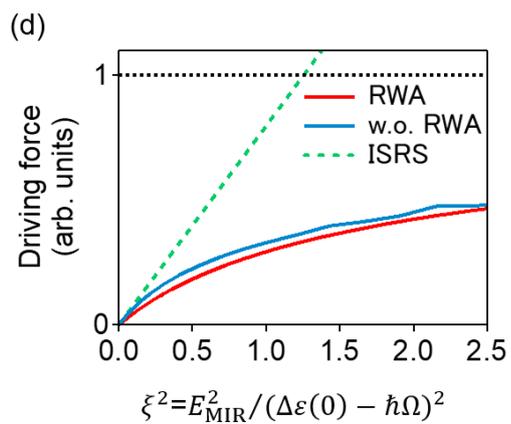

# Supplementary Information

K. Uchida[1], K. Nagai[1], N. Yoshikawa[1], and K. Tanaka[1,*]

[1] *Department of Physics, Graduate School of Science, Kyoto University, Kyoto, Kyoto 606-8502, Japan*

## 1. MIR-pump and NIR-probe experiment

We used 1 kHz Ti: Sapphire regenerative amplifier (35 fs, 7 mJ, 800nm) as a laser source. Figure S1 shows schematic of experimental setup. The part of laser output is used for generating MIR-pump pulses, and the other is used for probe pulses. MIR pulses are generated by difference frequency mixing of output from optical parametric amplifier system (Light Conversion TOPAS-C) in $AgGaS_2$ crystal. The signal and idler outputs are blocked by using a long-pass filter whose cutoff wavelength is 4 μm. MIR pulses were focused with an ZnSe lens (focal length 62.5 mm), and spot size at focal point was estimated to be around 60 μm in full-width half-maxima. The MIR polarization is set to be horizontal to optical table, and its amplitude is tuned by a pair of wire-grid polarizers. NIR probe pulse was focused with BK7 lens, and set to be nearly collinear to MIR-pump pulse by using D-shape mirror. The polarization angle of NIR pulses to optical table after reflecting D-shape mirror is set to be 45 degrees by using liquid crystal variable retarder (LCR). Transmitted NIR probe pulses passed through quarter wave plate and Wolaston prism, and then detected by balanced photodetector. In order to obtain high signal to noise ratio, we used a lock-in amplifier, synchronized to the modulation frequency of the MIR pulses (500 Hz). For the detection of high harmonic generation spectra, we used a spectrometer equipped with a Si-CCD camera. The background signal due to carrier generation is negligible compared with the oscillation signals as shown in

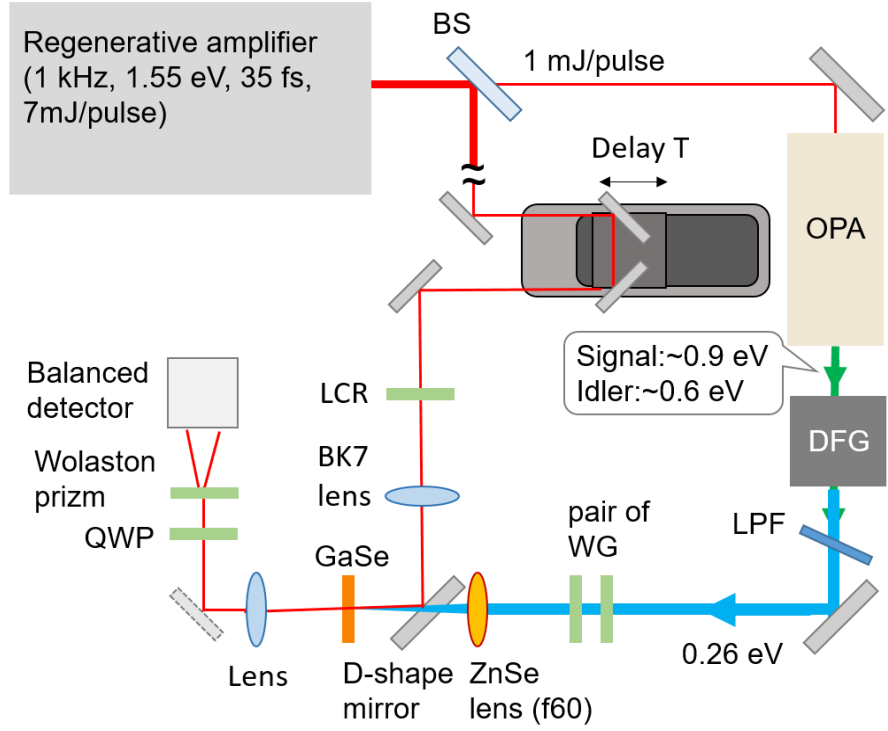

**Figure S1 Schematic of experimental setup.** BS: non-polarized beam splitter, LPF: low-pass filter, WG: wire-grid polarizer, LCR: liquid crystal retarder, QWP: quarter wave plate.

Fig 2b in the main text. This suggests that a small number of electron-hole pairs were generated by the MIR pulse excitation because the MIR photon energy (0.26 eV) is much smaller than the bandgap energy (2 eV). The maximum oscillation amplitude of the signal was as large as dI/I ~ $10^{-3}$, much larger than in conventional spontaneous Raman spectroscopy (~ $10^{-6}$).

## 2. Detection scheme of coherent phonon.

The in-plane Raman tensors of measured coherent phonon $A_{1g}$ and $E_{2g}$ (x or y) modes in the laboratory coordinate are given by[18]

$$\partial \chi / \partial Q_{A_{1g}} = \begin{pmatrix} a & \\ & a \end{pmatrix}, \tag{S1}$$

$$\partial\chi/\partial Q_{E_{2g}(x)} = \begin{pmatrix} -d\sin 2\theta & d\cos 2\theta \\ d\cos 2\theta & d\sin 2\theta \end{pmatrix}, \tag{S2}$$

$$\partial\chi/\partial Q_{E_{2g}(y)} = \begin{pmatrix} d\cos 2\theta & d\sin 2\theta \\ d\sin 2\theta & -d\cos 2\theta \end{pmatrix}, \tag{S3}$$

where $a$ and $d$ are the Raman coefficients, and θ is defined as the angle between horizontal axis and zigzag direction of GaSe crystal. In ISRS process, the driving force for phonon is given by[6,7]

$$F_i = \frac{1}{2}\frac{\partial \chi_{jk}}{\partial Q_i} E_j E_k. \tag{S4}$$

Here, $E_j$ is $j$-th (x or y) component of MIR electric field.

When we consider horizontally polarized MIR electric field, whose amplitude is given by $E_0$, driving force for each phonon mode is described by

$$F_{A_{1g}} = a/2\, E_0^2, \tag{S5}$$

$$F_{E_{2g}(x)} = d/2\, E_0^2 \sin(2\theta), \tag{S6}$$

$$F_{E_{2g}(y)} = d/2\, E_0^2 \cos(2\theta). \tag{S7}$$

Since we use MIR electric field along zigzag or armchair direction (θ = 0 or 90 degrees) in our experimental condition, we induced the $E_{2g}$ shearing motion along armchair direction ($E_{2g}$(y)). In the $E_{2g}$ mode, the direction of shearing motion depends on the incident MIR polarization. On the other hand, the absolute values of driving force are independent of the incident MIR polarization for both $A_{1g}$ and $E_{2g}$ modes. This result is inconsistent with our experimental result in non-perturbative regime: the saturation behavior of driving force depending on orientation angle in the $E_{2g}$ mode, indicating the breakdown of ISRS process.

The induced coherent phonon motions modulate the optical susceptibility of the sample in NIR region as follows:

$$\chi_{jk}(Q_i(t)) \approx \chi_{jk}(0) + \sum \frac{\partial \chi_{jk}}{\partial Q_i} Q_i(t). \tag{S8}$$

Here, we assumed that the optical susceptibility response to the phonon oscillation of $i$-th mode $Q_i(t)$ is instantaneous. When we consider $Q_i(t)=Q_i\cos(\Omega_i t)$, transmission change of NIR probe pulse in Fourier domain with sufficiently thin sample thickness is given by[7]

$$\Delta E_j^T(\omega, \tau) \approx i \frac{\omega l}{2cn} \frac{\partial \chi_{jk}}{\partial Q_i} Q_i \left( E_k^T(\omega - \Omega_i)e^{i\Omega_i \tau} + E_k^T(\omega + \Omega_i)e^{-i\Omega_i \tau} \right), \tag{S9}$$

where $c$ is the speed of light, $n$ and $l$ are refractive index and thickness of the sample, $\tau$ is the arrival time of probe pulse. The first and second terms respectively correspond to Stokes and anti-Stokes shifts in Raman scattering process.

For example, when the MIR polarization is along zigzag direction ($\theta = 0$ degree), and transmitted NIR electric field without pump is given by

$$E_x^{T0} = \frac{E_0}{\sqrt{2}} \delta(t-\tau)\cos\omega_0(t-\tau), \tag{S10}$$

$$E_x^{T0} = \frac{E_0}{\sqrt{2}} \delta(t-\tau)\cos\omega_0(t-\tau), \tag{S11}$$

where $\omega_0$ is the center frequency of the NIR probe, the resultant modulated transmitted NIR electric field is described as follows:

$$E_x^T(\omega,\tau) = \frac{E_0}{\sqrt{2}}\left\{1 + i\frac{\omega l}{2cn}\left(aQ_{A_{1g}}\cos\Omega_{A_{1g}}\tau + dQ_{E_{2g}}\cos\Omega_{E_{2g}}\tau\right)\right\}, \quad (S12)$$

$$E_y^T(\omega,\tau) = \frac{E_0}{\sqrt{2}}\left\{1 + i\frac{\omega l}{2cn}\left(aQ_{A_{1g}}\cos\Omega_{A_{1g}}\tau - dQ_{E_{2g}}\cos\Omega_{E_{2g}}\tau\right)\right\}, \quad (S13)$$

Note that when we consider the MIR polarization along armchair direction, we only need to change the sign of $d$. After passing through a quarter wave plate, where angle between horizontal axis and fast axis is defined as $\phi$, NIR probe electric field $E^D$ is given by

$$E_x^D = \frac{1}{\sqrt{2}}\{(1 + i\cos 2\phi)E_x^T + i\sin 2\phi\, E_y^T\}, \quad (S14)$$

$$E_y^D = \frac{1}{\sqrt{2}}\{(1 - i\cos 2\phi)E_y^T + i\sin 2\phi\, E_x^T\}. \quad (S15)$$

In our experimental setup, we detected the intensity difference between the horizontally and vertically polarized NIR probe intensities by using a Wolaston prism and a balanced photodetector. Hence, the resultant signal up to the 1st order of the Raman coefficients is given by

$$I_x^D(\omega,\tau) \propto 2 + \sin 4\phi + \frac{2\omega l}{cn}\sin 2\phi\, dQ_{E_{2g}}\cos\Omega_{E_{2g}}\tau, \quad (S16)$$

$$I_y^D(\omega,\tau) \propto 2 - \sin 4\phi - \frac{2\omega l}{cn}\sin 2\phi\, dQ_{E_{2g}}\cos\Omega_{E_{2g}}\tau. \quad (S17)$$

$$\Delta I(\omega,\tau)/I = \frac{1}{2}\sin 4\phi + \frac{\omega l}{cn}\sin 2\phi\, dQ_{E_{2g}}\cos\Omega_{E_{2g}}\tau. \quad (S18)$$

As a result, we can obtain the maximum $E_{2g}$ mode oscillation in the signal when we set $\phi = \pi/4$. The reason why the $A_{1g}$ mode cannot be obtained in Eq. (S18) is that the Stokes and anti-Stokes contributions cancel each other in this setup[7]. Thus, we need to make difference between $E_k^T(\omega - \Omega_{A1g})$ and $E_k^T(\omega + \Omega_{A1g})$ in Eq. (S9) to obtain finite signal from the $A_{1g}$ mode. In our experiment, polarization, amplitude, and phase depending on $\omega$, which are introduced through the usage of optics such as liquid crystal retarder, causes imbalance between $E_k^T(\omega - \Omega_{A1g})$ and $E_k^T(\omega + \Omega_{A1g})$, and enable us to detect both $A_{1g}$ and $E_{2g}$ mode at balanced point.

For example, we calculated the coherent phonon signals when $\phi$ is set to be $\pi/4$ by using NIR Gaussian pulse (FWHM: 35 fs) passing through an ideal half-wave plate and a LCR. Here, since the refractive indices of liquid crystal in the frequency region where NIR probe spectrum covers are almost independent of frequency for both ordinary and extra-ordinary rays, we consider that the Jones matrix $J$ for the variable liquid retarder can be written as follows:

$$J(\omega, \phi) = \begin{pmatrix} e^{i\pi\omega/2\omega_0} \cos^2(\frac{\pi}{8}) + e^{-i\pi\omega/\omega_0} \sin^2(\frac{\pi}{8}) & i\sin(\pi\omega/2\omega_0)/\sqrt{2} \\ i\sin(\pi\omega/2\omega_0)/\sqrt{2} & e^{-i\pi\omega/2\omega_0} \cos^2(\frac{\pi}{8}) + e^{i\pi\omega/\omega_0} \sin^2(\frac{\pi}{8}) \end{pmatrix}$$

(S19)

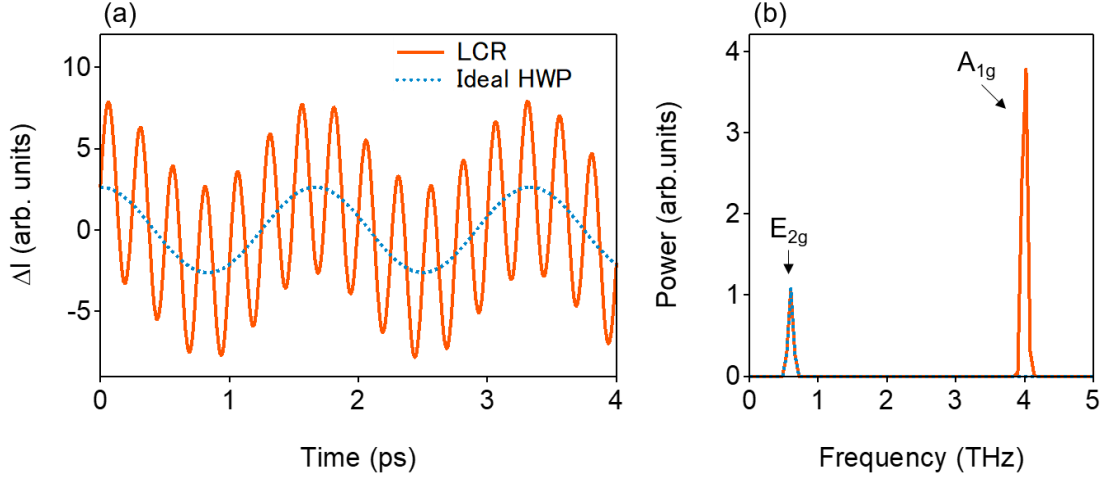

**Figure S2 Calculated pump-probe signal** (a) Temporal Oscillation signals by considering LCR (orange solid line) and ideal half-wave plate (blue dotted line). (b) The Fourier spectra of the signals corresponding to that in (a).

As a result, as shown in Fig. S2, both $A_{1g}$ and $E_{2g}$ modes can be detected by using LCR. Note that the absolute value of oscillation phase cannot be used as a probe of the phonon dynamics. This is because the phase of oscillation depends on the ratio between $E_k^T(\omega - \Omega_{A1g})$ and $E_k^T(\omega + \Omega_{A1g})$.

## 3. Derivation of driving force on phonon in non-perturbative regime.

To derive the driving force on phonon, we consider Hamiltonian of the system as follows:

$$H = H_e + H_{eN} + H_{eR} + H_N. \tag{S20}$$

Here, indices of e, N, eN, and eR correspond to electron, phonon, electron-phonon interaction, and light-electron interaction Hamiltonian. First, we consider only electron dynamics with a given phonon displacement $Q$ as follows:

$$i\hbar \frac{\partial}{\partial t} \phi_{i*}(Q,t) = \left(H_e + H_{eN}(Q) + H_{eR}(t)\right)\phi_{i*}(Q,t). \quad (S21)$$

Here, $\phi_{i*}(t)$ is the wavefunction of electron labeled by $i*$.

When we assume that $H_{eR}$ is periodic in time domain with periodicity of $2\pi/\Omega$ ($H_{eR}(t) = H_{eR}(t+2\pi/\Omega)$), we can employ the Floquet picture and apply Floquet theorem. In this case, the solution of Eq. (S21) can be written by[19]

$$\phi_{i*}(Q,t) = e^{-i\varepsilon_i^*(Q)t/\hbar}\varphi_{i*}(Q,t) = e^{-i\varepsilon_i^*(Q)t/\hbar}\sum_l \varphi_{i*,l}(Q)e^{-il\Omega t}, \quad (S22)$$

where $\varepsilon_i^*(Q)$ and $\varphi_{i*}(Q,t)$ is respectively called quasienergy and Floquet state, and $\varphi_{i*,l}(Q)$ is $l$-th Fourier component of Floquet state.

By using this result, the solution of Schrödinger equation $\Phi$ for total Hamiltonian in Eq. (S20) is given by

$$\Phi(Q,t) = \sum_{i*} C_{i*}(Q,t)\phi_{i*}(Q,t), \quad (S23)$$

where $C_{i*}(Q,t)$ is the coefficient for $i*$-th electronic state and reflects the dynamics of phonon. By substituting Eq. (S23), we obtain Schrödinger-like equation as follows:

$$i\hbar \frac{\partial}{\partial t} C_{i*}(Q,t) = \left(H_N + \varepsilon_i^*(Q)\right)C_{i*}(Q,t) +$$

$$\sum_j \langle \phi_{i*}(Q,t)|H_N|\phi_j(Q,t)\rangle C_j(Q,t). \quad (S24)$$

Here, by assuming Born-Oppenheimer approximation, we neglect the second term, i.e., diabatic transition between two Floquet states through phonon motion. As a results, the dynamics of phonon is described by

$$\Phi(Q,t) \approx C_i(Q,t)\phi_i(Q,t), \quad (S25)$$

$$i\hbar\frac{\partial}{\partial t}C_i(Q,t) = \left(H_N + \varepsilon_i^*(Q)\right)C_i(Q,t). \quad (S26)$$

In classical limit, the driving force for phonon induced by photon dressed state is given by

$$F_i = -\frac{\partial}{\partial Q}\varepsilon_i^*(Q)\Big|_{Q=0}. \quad (S27)$$

Thus, by calculating quasienergy depends on $Q$, we can derive driving force for phonon[20].

For simplicity, we employ two-level model as follows:

$$H_e + H_{eN}(Q) + H_{eR}(t) = \varepsilon_g(Q)|g\rangle\langle g| + \varepsilon_e(Q)|e\rangle\langle e| + $$
$$dE_{MIR}(t)|g\rangle\langle e| + d^*E_{MIR}(t)|e\rangle\langle g|. \quad (S28)$$

Here, $\varepsilon_g(Q)$ ($\varepsilon_e(Q)$) is the electronic energy of the ground (excited) state with adiabatic potentials with respect to a normal coordinate $Q$, $d=\langle g|\hat{d}|e\rangle$ is the transition dipole moment between ground and excited states, and $E_{MIR}(t)$ is the temporal profile of the MIR electric field. We regard MIR electric field as continuous wave ($E_{MIR}(t) = E_{MIR}\cos\Omega t$), and apply the rotating-wave approximation (RWA). This enable us to analytically derive the quasienergy depending on the MIR electric field strength: so-called Rabi splitting[21].

$$\varepsilon_g^*(Q) = \varepsilon_g(Q) + \frac{(\Delta\varepsilon(Q)-\hbar\Omega)}{2} - \frac{\sqrt{d^2E_{MIR}^2+(\Delta\varepsilon(Q)-\hbar\Omega)^2}}{2}, \quad (S29)$$

$$\varepsilon_e^*(Q) = \varepsilon_e(Q) - \frac{(\Delta\varepsilon(Q)-\hbar\Omega)}{2} + \frac{\sqrt{d^2E_{MIR}^2+(\Delta\varepsilon(Q)-\hbar\Omega)^2}}{2}. \quad (S30)$$

Here, $\Delta\varepsilon = \varepsilon_e - \varepsilon_g$ is the transition energy. We assume electronic state stays dressed ground state g*. In this case, the driving force for phonon can be written by

$$F = -\frac{\partial}{\partial Q}\varepsilon_g^*(Q)\bigg|_{Q=0} = -\frac{1}{2}\frac{\partial}{\partial Q}\varepsilon_e(Q)\bigg|_{Q=0}\left(1 - \frac{1}{\sqrt{1+\xi^2}}\right), \quad (S31)$$

$$\xi = \frac{dE_{MIR}}{\Delta\varepsilon(0) - \hbar\Omega}, \quad (S32)$$

where $\xi$ is known as a parameter for the perturbation expansion in nonlinear optics[22]. In our calculation, we use $\Delta\varepsilon(0) = 2.0$ eV corresponding to bandgap energy in GaSe, and $\hbar\Omega = 0.26$ eV. When $\xi$ is much smaller than one, where perturbative nonlinear optics is valid, driving force becomes

$$F \approx -\frac{1}{2}\frac{\partial}{\partial Q}\varepsilon_e(Q)\bigg|_{Q=0}\xi^2. \quad (S33)$$

This result is almost identical to that of ISRS process for two level system, which is given by

$$F = -\frac{1}{2}\frac{\partial}{\partial Q}\varepsilon_e(Q)\bigg|_{Q=0}\xi^2 - \frac{1}{2}\frac{\partial}{\partial Q}\varepsilon_e(Q)\bigg|_{Q=0}\left(\frac{dE_{MIR}}{\Delta\varepsilon(0)+\hbar\Omega}\right)^2. \quad (S34)$$

Note that the additional second term is the contribution from counter-rotating electric field, which is neglected in the RWA.

On the other hand, when $\xi$ becomes unity, where non-perturbative nonlinear optical phenomena such as deviation from powers law in HHG can be observed, the driving force starts to deviate from Eq. (S33), and then, gradually approach the upper limit of driving force with the increase of $\xi$ as follows:

$$\lim_{\xi\to\infty} F = -\frac{1}{2}\frac{\partial}{\partial Q}\varepsilon_e(Q)\bigg|_{Q=0}. \quad (S35)$$

This tendency has good agreement with our experimental result, i.e., phonon amplitude saturation in accordance with the breakdown of powers law in HHG.

**4. Driving force on phonon without the RWA.**

In our experiment setup, MIR photon energy (0.26 eV) is much smaller than the bandgap energy of GaSe (~2 eV). In this situation, we cannot neglect the contribution from counter-rotating field, and the result using the RWA is not necessarily valid. Hence, we perform the numerical calculation of the driving force without the RWA to check the validity of our calculation in the main text.

In the Floquet theory, time dependent Schrodinger equation can be regarded as the static problem as follows:

$$\sum_m (H_{l-m} - m\Omega\delta_{lm}) \varphi_{i*,m}(t) = \varepsilon_i^* \varphi_{i*,l}, \tag{S36}$$

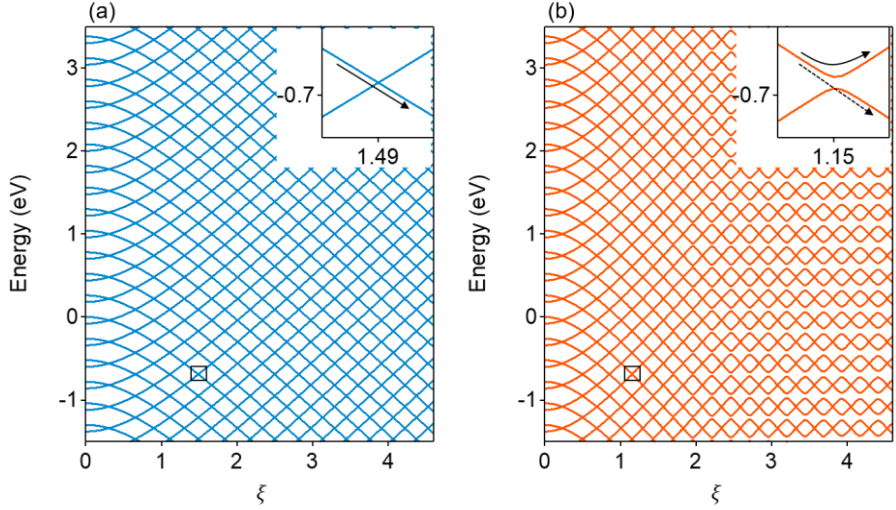

**Figure S3 Quasi-energy spectra as a function of ξ (a) with and (b) without RWA.** The insets show the extended figures, whose plotted regions are indicated by black squares, respectively. With RWA, there is no anti-crossing behavior, and electron dynamics with the increase of electric field can be described by the solid arrow in the inset of (a). Without RWA, there are anti-crossing behaviors, and the solid (dashed) arrow in the inset of (b) indicates the dynamics of electron with respect to the increase of electric field amplitude in the adiabatic (diabatic) limit.

$$\left(H_e + H_{eN}(Q) + H_{eR}(t)\right) = \sum_l e^{-il\Omega t} H_l. \tag{S37}$$

In the calculation, we numerically solve Eq. (S36) with Hamiltonian in Eq. (S28) without the RWA by considering Fourier component of $-160 \leq l \leq 160$. Figure S3 shows the resultant quasienergy spectra as a function of $\xi$ (a) with and (b) without the RWA. By considering counter-rotating electric field, there are two effects on quasienergy spectra. One is Bloch-Siegert shift, which can be described by $d^2 E_{MIR}^2/(\Delta\varepsilon + \hbar\Omega)$ in the lowest perturbation limit[23]. This causes a slight additional red (blue) shift in the quasienergy of "dressed" ground (excited) state. The other is anti-crossing behavior at the crossing point of the "dressed" ground and excited sidebands (the inset of Fig. S3 (b)). When the envelope of MIR electric field varies sufficiently slow (adiabatic limit), the flopping from the "dressed" ground state to excited state with the RWA is induced by the counter-rotating electric field. On the other hand, when the envelope varies fast (diabatic limit), electronic state stays "dressed" ground state with RWA. Since MIR pulse width of 60 fs (FWHM in intensity) is comparable to the period of MIR light of 16 fs in our experimental setup, we consider that our experimental condition is diabatic limit.

Figure S4 (a) shows the driving force for phonon as a function $\xi^2$ in diabatic limit. When $\xi \ll 1$, the driving force is identical to that of ISRS process described in Eq. (S34). Due to the effect of Bloch-Siegert shift, the driving force without the RWA is slight stronger than that with the RWA, but the saturation behavior is well reproduced.

On the other hand, when we assume adiabatic limit, there are sudden increases and decreases of driving force across the upper limit derived with the RWA (Fig. S4(b)). The blue and red dashed lines in Fig. S4(b) show the driving force in the "dressed" ground and excited states with the RWA. This clearly indicates that the sudden increase and decrease of the driving force is caused by the flopping from the "dressed" ground to excited states with the RWA. Although the driving force exceeds the upper limit with the RWA, this is now limited by the driving force in the "dressed" excited state with the

RWA. These results suggest that we can tune the driving force for phonon through the

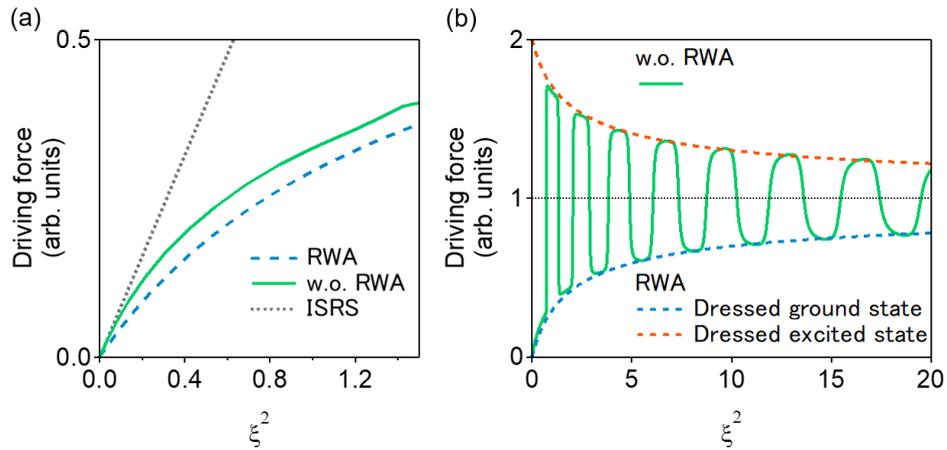

**Figure S4 Driving force on phonon.** (a) driving force on phonon as a function of $\xi^2$ in diabatic limit. Dashed blue line indicates the analytical result with RWA. Green solid line indicates numerical result without RWA. Gray dotted line indicate driving force in ISRS process. (b) driving force on phonon in adiabatic limit. Blue and orange dashed line indicate driving force of "dressed" ground and excited states with RWA, respectively. Green solid line indicates that of "dressed" ground state without RWA. Black dotted line indicates the upper limit of the driving force with RWA.